\documentstyle[12pt]{article}

\def\be{\begin{equation}}
\def\ee{\end{equation}}

\def\bea{\begin{eqnarray}}
\def\eea{\end{eqnarray}}

\begin{document}

\baselineskip15pt

\noindent

\title{Dynamical equivalence, commutation relations and noncommutative geometry\thanks{Dedicated to the 75th birthday of Prof.~Jan Lopuszanski} \thanks{Supported by the Humboldt Foundation} }

\author{P.C.~Stichel\\
An der Krebskuhle 21, D-33619 Bielefeld, Germany\thanks{e-mail: hanne@physik.uni-bielefeld.de}}

\date{}

\maketitle

\begin{abstract}
We revisit Wigner's question about the admissible commutation relations for coordinate and velocity operators given their equations of motion (EOM). In more general terms we want to consider the question of how to quantize dynamically equivalent Hamiltonian structures. A unique answer can presumably be given in those cases, where we have a dynamical symmetry. In this case arbitrary deformations of the symmetry algebra should be dynamically equivalent. We illustrate this for the linear as well as the singular $1d$-oscillator. In the case of nonlinear EOM quantum corrections have to be taken into account. We present some examples thereof.
New phenomena arise in case of more then one degree of freedom, where sometimes the interaction can be described either by the Hamiltonian or by nonstandard commutation relations. This may induce a noncommutative geometry (for example the $2d$-oscillator in a constant magnetic field). Also some related results from nonrelativistic quantum field theory applied to solid state physics are briefly discussed within this framework.
\end{abstract}

\section{Introduction}

It is well known, that the Lagrangean leading to a given description of a classical mechanical system is not unique. To be more specific, we have to ask for the set of all Lagrange functions, whose Euler-Lagrange equations have the same solutions in configuration space. Those 
Lagrangeans are called $s$-equivalent. The task of finding them has been solved completely for systems with one degree of freedom in terms of one arbitrary positive function [1], while an extensive discussion for two degrees of freedom has been given by Douglas [2]. As classical
dynamics is described completely by trajectories in configurations space, $s$-equivalent Lagrangeans are dynamically equivalent. Dynamical equivalence my also be expressed in terms of a set of equivalent Hamiltonian structures $\{ (\omega, H)\}$ where $\omega$ denotes a symplectic structure (fundamental Poisson brackets) and $H$ a Hamilton function. In turning to quantum mechanics, a fundamental question arises:

\noindent
How to quantize dynamically equivalent Hamiltonian structures?

\noindent
One may ask also the subquestion: What are, given the equations of motion (EOM) for coordinate and velocity operators, the admissible commutation relations between them? This question was first asked by Wigner [3] for the $1d$-harmonic oscillator in the framework of the standard Hamiltonian. He got a whole set of solutions, which are characterized by one real parameter [3]. His solutions are equivalent to the parabose algebra, as has been shown by Palev [4a].

Proceeding from $s$-equivalent Lagrangeans, Wigner's question was first considered by Okubo [5] for some unconventional examples. A general treatment of this question for quantum mechanical systems living on a finite dimensional Hilbert space as well as for the $1d$-oscillator (infinite dimensional Hilbert space) has been given very recently by Man'ko, Marmo, Sudarshan and Zaccaria [6]. For reasons of completeness we will take up the discussion of the $1d$-oscillator again in the present paper. In particular we will discuss parabosons and different cases of $q$-deformations within a unique framework of nonlinear deformations of the oscillator algebra. In addition we examine representations for some simple nonlinear deformations and discuss the essential difference between the classical and quantum mechanical formalism for general deformations. 

The harmonic oscillator is an exceptional case insofar, as the EOM are linear. They are identical in both classical mechanics and quantum mechanics. Therefore, in passing from classical Poisson brackets to commutators by means of Dirac's recipe we have no difficulties. But the situation becomes worse if nonlinear observables besides the Hamiltonian are involved. Then we are confronted with the inconsistency of Dirac's rule [7]. In those cases quantum corrections appear either in the EOM or in some observables. This will be demonstrated for the singular oscillator, for some power potentials and for spherically symmetric potentials. 

In this context we will neither discuss modern treatments of the quantization or dequantization problem (cp [7], [8], [9]) nor their difficulties [10]. It is the aim of our
paper, to consider our fundamental question by means of some simple but important physical examples, but not to discuss it in terms of a general mathematical framework.

Another very interesting question is the prescription of the interaction not in terms of a Hamiltonian but in terms of nonstandard commutation relations. We will demonstrate this for the example of a charged particle moving in a constant magnetic field in a plane. The resulting
nonstandard commutation relations describe a noncommutative geometry. It is an exciting topic to extend this question to quantum field theory. We give a brief account of some related results in solid state physics.

The paper is organized as follows: In Sec.~2 we treat systems with one degree of freedom, divided into Hamiltonian mechanics with the most general symplectic form, quantum mechanics of the nonlinear deformed linear and singular oscillator and the case of a general potential. Sec.~3 is devoted to the movement of a charged particle in a constant magnetic field in a plane. In Sec.~4 spherically symmetric potentials are revisited. 
Sec.~5 contains some remarks on examples from nonrelativistic quantum field theory. In Sec.~6 we close with some final remarks including open questions. 

\section{Systems with one degree of freedom}

In this section we study dynamical equivalence for the motion of either one particle in an external field or for the relative motion of two particles in $1d$-space within the framework of classical or quantum mechancis.

\subsection{${\bf 1d}$ -- classical mechanics}

For reasons of simplicity we consider Newton's EOM for a conservative force only
\be
\ddot{x} = - V^\prime (x) \ .
\ee
The corresponding description in the standard formulation of the canonical formalism is given by the Hamilton function
\be
H(u,x) = \frac{u^2}{2} + V(x)
\ee
leading by means of the symplectic structure $\omega_0$ for the independent variables

\noindent
$(y_1,y_2) = (x,u)$ with
\be
(\omega_0)_{ij} = \epsilon_{ij}
\ee
to the canonical EOM
\begin{eqnarray}
\dot{x} &=& \{ x,H \}_{\omega_0} = u \\
\dot{u} &=& \{ u,H \}_{\omega_0} = -V^\prime (x)
\end{eqnarray}
and therefore to Newton's EOM (1) in $x$-space where we define the Poisson bracket $\{ \cdot , \cdot \}_\omega$ for an arbitrary symplectic structure $\omega$ as usual\footnote{We use the summation convention for repeated indices.}
\be
\{ A,B \}_\omega : = \frac{\partial A}{\partial y_i} \omega_{ij} 
\frac{\partial B}{\partial y_j} \ .
\ee
Now we ask for the most general Hamiltonian structure $(\omega, \tilde{H})$ with 
\be
\frac{\partial\omega}{\partial t} = 0
\ee
which preserves the EOM (4), (5), i.e.
\begin{eqnarray}
\dot{x} &=& \{ x,\tilde{H} \}_\omega = u \\
\dot{u} &=& \{ u,\tilde{H} \}_\omega = - V^\prime (x) 
\end{eqnarray}
Such a structure $(\omega, \tilde{H})$ we call dynamically equivalent to $(\omega_0, H)$. According to Leubner and Marte [11] $\omega$ is a conserved quantity, which due to (7) has to be a function of $H$ only. Therefore $\omega$ may be expressed in terms of an arbitrary nonvanishing function $\sigma (z)$ (which we choose to be positive), such that
\be
\omega = \left( \begin{array}{cc}
0  & 1/\sigma (H) \\
-1/\sigma (H)  & 0 \end{array} \right) 
\ee
It is easily seen, that for a given $\sigma$ the Hamiltonian structure $(\omega, \tilde{H})$ with
\be
\tilde{H} : = \int^H dz\sigma (z)
\ee
satisfies the EOM (8), (9).
\smallskip

\noindent
\underline{Remark:} The foregoing results my be derived also within the Lagrangean framework [1].

\subsection{Quantum mechanics of the deformed \protect$1d$-oscillator}

The EOM in configuration space for the linear oscillator has the form
\be
\ddot{x} + x = 0
\ee
which may be written as a system of two first order equations
\be
\dot{x} = u, \qquad \dot{u} = -x \ .
\ee
If we introduce raising and lowering operators as usual
\be
a^\dagger : = \frac{1}{\sqrt{2}} (x - iu), ~~~~~~~a : = \frac{1}{\sqrt{2}} (x + iu)
\ee
the EOM (13) takes the form
\be
\dot{a} = - i a, \qquad \dot{a}^\dagger  = i a^\dagger\ .
\ee
Due to (15) the operator $N$
\be
N : = a^\dagger a
\ee
as well as the commutator $[ a,a^\dagger ]$ are conserved quantities.

\noindent
We conclude, as in classical mechanics (cp. section 2.1), that the commutator may be expressed by a positive function $\sigma$ which has to be a function of $N$ only
\be
[ a,a^\dagger ] = \frac{1}{\sigma(N)}\ .
\ee
For the standard description of the linear oscillator we take $\sigma = 1$. The generalization (17) defines an arbitrary deformation of the usual oscillator algebra. Now we ask for a new number operator $K$, related to the new Hamiltonian $\tilde{H}$ through $\tilde{H} = K + 1/2$, which preserves the EOM (15)
\begin{eqnarray}
\left[ K , a \right] &=& -a \\
\left[ K , a^\dagger \right] &=& a^\dagger 
\end{eqnarray}

\noindent
Taken as a function of $N$ our new number operator $K$ satisfies the functional equation
\be
K\left( N+\frac{1}{\sigma(N)} \right) = 1 + K(N)\ .
\ee
The solution of (20) is supposed to be unique up to a constant. 

\noindent
We prove (20) by starting with
\be
N a^\dagger - a^\dagger (N + \frac{1}{\sigma(N)} = 0
\ee
which follows from (16), (17). Therefore, for each holomorphic function $f$ we obtain the relation
\be
f(N) a^\dagger - a^\dagger f\left( N+ \frac{1}{\sigma(N)} \right) = 0\ .
\ee
By identifying $f(N)$ with $K(N)$ and using (19) we obtain immediately (20). There exists an useful alternative formulation of (20): Suppose the commutator $[a,a^\dagger]$ is given instead 
of by (17) in terms of a positive function $\varphi(K)$
\be
[ a,a^\dagger ] = \varphi(K)
\ee
and we have an implicit definition of $K$ by means of a positive function $F$
\be
N = F(K)\ .
\ee
Then, by applying $F$ to (20) we obtain the relation
\be
\varphi (K) = F(K+1) -F(K)
\ee
between the functions $F$ and $\varphi$ (cp. [12], [13]).

\noindent
There is a corollary to relation (25). Let us define the $q$-commutator (quommutator) $[ a,a^\dagger ]_q$ by\footnote{We don't consider complex valued q's in this paper.}
\be
[ a,a^\dagger ]_q : = a a^\dagger - q a^\dagger a,\qquad  q \in {\bf R}^1\ .
\ee
Then the commutator (23) with $\varphi$ given by (25) and the quommutator
\be
[ a, a^\dagger ]_q = \psi (K)
\ee
are equivalent if
\be
\psi (K) = F (K+1) - qF(K)\ .
\ee

\bigskip
We observe, that our quantum mechanical relation (25) looks different from the corresponding relation in classical mechancis, which, due to (11) has the form
\be
\varphi (z) = F^\prime (z)
\ee
with $z \in {\bf R}^1_+$. But as opposed to (29) our $K$ in (25) is not a continuous variable but an operator taking discrete values $n \in {\bf N}$ on a $1d$-lattice space. In order to compare (25) with (29) we have to introduce a differential calculus on this lattice space (cp. [14]). The simplest way to do this is to introduce an operator valued differential $d g(K)$ for an arbitrary holomorphic function $g(z)$. We define
\be
d g (K) : = [a, g (K)]\ .
\ee

\bigskip
\noindent
\underline{Remarks} on the definition (30):

\medskip
\noindent
1) It respects the Leibniz rule $d(gh) = (dg) h + g dh$.

\medskip
\noindent
2) By means of Dirac's rule the dequantization looks as follows
\[
dg(K) \to i\hbar \{ a,g(K)\}_\omega = g^\prime (K) \hbar a\ .
\]

\medskip
\noindent
If we specify $g(z) = z$ we get due to (18)
\be
dK = a\ .
\ee

\medskip
\noindent
With that and using (18) again we finally obtain for (30)
\be
dg (K) = (g (K+1) - g(K)) dK\ .
\ee
This is a differential calculus in noncommutative geometry (cp. [14]), because $dK$ and $K$ don't commute. On the contrary we find
\[
[ dK,K ] = dK\ .
\]
Let us define the left partial derivative of $g(K)$ by [14]
\be
dg(K) = (\partial_+ g) dK
\ee
then we recognize $\varphi(K)$ in (25) as just this derivative of $F(K)$.

\bigskip
\noindent
\underline{Remark:} Instead of (30) we could have defined another operator valued differential 
\[
\tilde{d} g(K) : = [g(K), a^\dagger]
\]
leading finally to the identification of $\varphi (K)$ as the right partial derivative of $F(K)$ (with respect to the differential $\tilde{d}$). 

\smallskip
\noindent
We conclude, that agreement between the quantum mechanical and classical expressions (25) and (29) respectively may be obtained, if we use appropriate differential calculi in both cases.

\bigskip
The foregoing description of deformed oscillators yields a general and unique framework covering all the results known already. We want to illustrate this with some typical examples.

\bigskip
\noindent
{\bf (1)} The algebra of the Arik-Coon $q$-oscillator [15]
\be
[ a,a^\dagger ]_q = 1
\ee
is equivalent to (cp. [13])
\be
[ a,a^\dagger ] = q^K \ .
\ee
This follows immediately from $\psi = 1$ and (28) leading to
\be
F(K) = \frac{q^K -1}{q-1}\ .
\ee
If we take the inverse of (36) we get
\be
K(N) = \frac{1}{lnq} ln (1+(q-1)N)
\ee
which is obviously the solution of (20) with
\[
\frac{1}{\sigma (N)} = 1 + (q-1)N
\]
obtained from (34).

\bigskip
\noindent
{\bf (2)} The algebra of the Biedenharn-Macfarlane $q$-oscillator [16]
\be
[ a,a^\dagger ]_q = q^{-K}
\ee
is equivalent to
\be
[ a,a^\dagger ] = F(K+1) - F(K)
\ee
with
\be
F(K) = \frac{q^K - q^{-K}}{q-q^{-1}} \ .
\ee
By inverting (40) we obtain with $q = e^\lambda$
\be
K(N) = \frac{1}{\lambda} ln (N\sinh \lambda + (N^2 \sin^2 h\lambda +1)^{1/2})\ .
\ee
The corresponding $\sigma(N)$ is given by 
\[
1/\sigma(N) = N (\cosh \lambda -1) + (N^2 \sin^2 h
\lambda + 1)^{1/2}\ .
\]

\medskip
\bigskip
Examples (1) and (2) show the equivalence between commutators and quommutators.

\medskip
\noindent
Systematic constructions of such an equivalence is given in [13] by means of a recursive procedure in Fock-space.

\medskip
\bigskip
\noindent
{\bf (3)} The parabose oscillator of order $p$ is defined by\footnote{the symbol $\{ \cdot, \cdot \}$ denotes the anticommutator} (cp. [b4])
\be
K = \frac{1}{2} \{ a^\dagger, a\} - p/2, \qquad p \in {\bf N}
\ee
where the vacuum has to satisfy the relations
\be
a|0> = 0 \quad \mbox{and} \quad a a^\dagger | 0 > = p | 0 >\ .
\ee
In order to find the function $\varphi$ we rewrite (42) as
\be
K = F(K) + \frac{1}{2} \varphi (K) - p/2\ .
\ee
If we eliminate $F(K)$ in (25) by means of (44) we obtain an equation for $\varphi$
\[
\frac{1}{2} (\varphi (K) + \varphi (K+1)) = 1
\]
which has the solution
\be
\varphi (K) = 1 + \alpha (-1)^K, \qquad \alpha \in {\bf R}^1\ .
\ee
Finally by using (43) we obtain
\be
\alpha = p -1 \ .
\ee
This result has already been obtained in [4] by means of a more complicated procedure.

\medskip
\bigskip
The representation of the deformed oscillator algebra (18), (19), (23), in particular the spectrum of the Hamiltonian, depends on the deformation given by the function $\varphi$ (cp.~[17]). In order to illustrate this statement, let us consider a very simple example of a family of deformations 
\be
\varphi_q (K) = q + 2(q-1) K, \qquad q \in {\bf R}^1
\ee
which interpolates between oscillator algebra $(q=1)$, $SU(2)$ $(q=0)$ and $SU(1,1)$ $(q=2)$. But if for $q \notin 1$ we shift the scale of $K$
\be
K = K^\prime - \frac{q}{2(q-1)}
\ee
and rescale $a$ simultaneously
\be
a = | q-1 |^{1/2} a^\prime
\ee
we arrive for $q > 1$ ($q < 1$) at the $SU(1,1)$ ($SU(2)$) algebra respectively
\be
[ K^\prime, a^\prime ] = - a^\prime, \qquad [ K^\prime, a^{\prime \dagger} ] = a^{\prime \dagger},\qquad [a^\prime , a^{\prime\dagger}] = \pm 2K^\prime
\ee
where the upper (lower) sign has to be taken for $q > 1 (q < 1)$.

\noindent
We conclude that the limit $q\to 1$ in (47) is globally discontinuous.

\noindent
But this is not the whole story. The deformed oscillator algebra has the Casimir operator [18]
\be
C = N - F(K) \ .
\ee
But in eq. (24) we have indentified $F(K)$ with $N$ on our state space. Therefore, we can realize only those irreducible representations of our deformed algebra, which exhibit the value zero for $C$ corresponding to the value zero for the Casimir operator $C^\prime$
\be
C^\prime = N^\prime \mp K^\prime (K^\prime -1)
\ee
of the algebra (50). The latter may be seen as follows: Define for $q \not= 1$
\be
F_q (K) = K + (q-1) K^2 + \frac{1}{4(q-1)} - \frac{q-1}{4}
\ee
leading due to (25) to $\varphi_q (K)$ in (47). With the shift $K\to K^\prime$ we obtain
\be
F_q (K) = (q-1) K^\prime (K^\prime -1)
\ee
and therefore
\be
C = | 1-q | C^\prime \ .
\ee
The value zero for $C^\prime$ corresponds for $SU(2)$ to a singlet state -- the Hilbert space $\cal H$ consists of one state only! 

\bigskip
\noindent
\underline{Remark:} For more general deformations, $\cal H$ might consist of a finite number of states [18].

\bigskip
\noindent
For the $SU(1,1)$ case, a vanishing Casimir $C^\prime$ is degenerate: $\cal H$ is the direct sum of a singlet state $| 0 >$ and an infinite dimensional representation with $|1>$ as the cyclic vector. 

\noindent
Here we denoted by $|h>$ the eigenstates of $K^\prime$
\[
K^\prime | h > = h | h > \ .
\]
In both cases $F_q(K)$ is a nonnegative operator on $\cal H$ as required.

\bigskip
Only for reasons of completeness we note, that the eigenstates of $K$ in configuration or momentum space differ from the undeformed case (cp. [19], [20]). In particular, deformations of the oscillator algebra change Heisenberg's uncertainty relation. In certain cases a minimal length uncertainty shows up (cp. [21]). An example for that is given by the Arik-Coon oscillator.

\subsection{The singular oscillator}

It is our aim to demonstrate in this subsection, that $s$-equivalence may be destroyed in the quantum case if we require the validity of a deformed symmetry algebra. As an example we consider the $1d$-singular oscillator, defined by the potential (cp. [22])
\be
V(x) = \frac{x^2}{2} + \frac{b}{x^2}, \qquad b > 0
\ee
on the half-line $x \ge 0$.

\bigskip
\medskip
\noindent
\underline{Classical mechanics}

\medskip
\noindent
The standard canonical description shows a dynamical $SU(1,1)$-symmetry. We describe this symmetry in terms of the standard Hamiltonian
\be
H = \frac{u^2}{2} + V(x)
\ee
and
\be
K_\pm : = \frac{1}{2} \left( \frac{1}{2} ( x \mp iu)^2 - \frac{b}{x^2}\right)
\ee
as follows 
\begin{eqnarray}
\{ K_\pm, K_0\}_{\omega_0} &=& \pm i K_\pm\\
\{ K_-, K_+ \}_{\omega_0} &=& - 2 i K_0
\end{eqnarray}
with $K_0 : = H/2$.

\noindent
The EOM may be written either in the form (from (59))
\be
\frac{d}{dt} K_\pm = \pm 2 i K_\pm
\ee
or in the standard form
\begin{eqnarray}
\dot{x} &=& u \nonumber\\
\dot{u} &=& - (x-\frac{2b}{x^3}) \ .
\end{eqnarray}
It can easily be seen, that (61) and (62) are $s$-equivalent, because they are connected by a nonsingular linear transformation.

\noindent
According to the results presented in Sec. 2.1, the singular oscillator may be described in terms of a general Hamiltonian structure ($\omega, \tilde{H}$), which preserves the EOM in both forms. Accordingly we obtain a deformed $SU(1,1)$
\begin{eqnarray}
\{ K_\pm, K\}_\omega &=& \pm i K_\pm\\
\{ K_- , K_+\}_\omega &=& - \frac{iH}{\sigma (H)}
\end{eqnarray}
instead of (59), (60). We defined $K : = \tilde{H}/2$. Because $K$ is a function of $H$ due to (11), the r.h.s. of (64) may be expressed also in terms of $K$
\be
\{ K_- , K_+ \}_\omega = - i \psi (K) \ .
\ee

\bigskip
\bigskip
\noindent
\underline{Quantum mechanics}

\medskip
\noindent
In passing over from classical to quantum mechanics we want to keep the deformed $SU(1,1)$ algebra. Therefore, we don't apply Dirac's quantization recipe to the EOM of (62) but rather to (63). Our deformed $SU(1,1)$ then takes the form
\begin{eqnarray}
\left[ K, K_\pm \right] &=& \pm K_\pm \\
\left[ K_-, K_+ \right] &=& \psi (K)  \ .
\end{eqnarray}
But (66) is inconsistent with the quantization of the EOM (62) written in terms of Poisson brackets. By means of Dirac's rule we would obtain 
\be
i [\tilde{H},x] = u, \qquad i [\tilde{H},u] = - x + \frac{2b}{x^3}\ .
\ee
But let us now calculate the commutator $[ K,K_\pm ]$ by using (68) with (58). In this way we obtain
\be
[ K,K_\pm ] = \pm K_\pm + \frac{ib}{2} \left[ \frac{1}{x^2}, \frac{1}{x} [u,x] \frac{1}{x} \right]
\ee
instead of (66). If the commutator $[u,x]$ would have been a function of $x$ only, both expressions would have coincided. But such an exclusive $x$-dependent of $[u,x]$ is for a generic deformation in disagreement with (67).

\noindent
Because of $[ u,x ] = 0(\hbar)$ the additional term in (69) clearly is a quantum correction. It results from the well known fact, that the simultaneous application of Dirac's rule to different Poisson brackets is in general inconsistent (cp. [7]). 

\bigskip
\noindent
\underline{Problem:} What are the quantum corrections to (68) if we start with (66)?

\subsection{Arbitrary potentials \protect$V(x)$}

Not much can be said for an arbitrary potential $V(x)$ which differs from the harmonic one.

\noindent
Suppose we have the quantum analogon to the classical Hamiltonian structure $(\omega, \tilde{H})$ such , that corresponding to (8), (9) our EOM look as follows
\begin{eqnarray}
\frac{i}{\hbar} \left[ \tilde{H},x \right] &=& u\\
\frac{i}{\hbar} \left[ \tilde{H},u \right] &=& - V^\prime(x)\ .
\end{eqnarray}
By means of Jacobi's identity we conclude, that the commutator $[u,x]$ is conserved
\[
[\tilde{H}, [u,x]] = 0\ .
\]
But we are not able to express $[u,x]$ in terms of $H$ as in the case of the classical Poisson bracket.
With the exception of the oscillator or a constant force $H$ is not a conserved quantity if $[u,x]$ differs from a $c$-number. Therefore we must describe $[u,x]$ in terms of $\tilde H$ from the very beginning
\be
[u,x] = \frac{\hbar}{i} \varphi (\tilde{H})\ .
\ee

\bigskip
\smallskip
\noindent
\underline{Remark:} In simple cases for $\varphi$ and $V(x)$ we may construct a conserved extension $\hat{H}$ of $H$. For example consider the deformation
\be
\varphi (\tilde{H}) = 1 + \alpha \tilde{H}, \qquad \alpha \in {\bf R}^1
\ee
where $\alpha$ has the dimension of inverse energy, together with a power potential
\[
V(x) = \lambda x^n\ , \qquad n = 3, 4\ .
\]

\noindent
Then we obtain by straightforward calculation
\be
\hat{H}_3 = H_3 - \frac{\alpha \lambda}{2} \hbar^2 x
\ee
and
\be 
\hat{H}_4 = H_4 - \alpha \lambda \hbar^2 x^2
\ee
where we defined $H_n : = \frac{u^2}{2} + \lambda x^n$.

\noindent
In the limit $\alpha \to 0$ or $\hbar \to 0$ respectively, we obtain the standard result. The additional terms in (74), (75) are again quantum corrections.

\section{Systems with two degrees of freedom}

We are not going to generalize the consideration of Sec.~2 to two dimensions, but concentrate on those phenomena, which are typical for the $2d$-case. In particular we will consider a charged particle (charge $e$) moving in a constant magnetic field perpendicular to the plane of motion with (or without) an additional harmonic potential.

\noindent
In the presence of a harmonic potential the EOM in configuration space looks as follows
\be
\ddot{x}_i = - \omega^2 x_i + \kappa \epsilon_{ij}\dot{x}_j
\ee
with $\kappa : = \frac{e}{c} B$.

\noindent
The EOM (76) are linear equations. Therefore no problems arise with the quantization procedure and we may consider the quantum mechanical formulae from the beginning. First we consider the standard description, which is given by the Hamiltonian
\be
H_1 = \frac{p_i^2}{2} + \frac{\tilde{\omega}^2}{2} x^2_i - \frac{\kappa}{2} J
\ee
with angular momentum
\be
J : = \epsilon_{ij} x_i p_j
\ee
and shifted frequency 
\be
\tilde{\omega}^2 : = \omega^2 + (\kappa/2)^2\ .
\ee
By means of the canonical commutation relations
\begin{eqnarray}
\left[ p_i, x_j\right] &=& \frac{1}{i} \delta_{ij}\\
\left[ x_i, x_j\right] &=& 0\\
\left[ p_i, p_j\right] &=& 0
\end{eqnarray}
the EOM in phase space lead as usual to (76).

\noindent
As a first alternative we study the case of noncommuting space variables
\be
\left[ x_i, x_j \right] = i \frac{\kappa}{\omega^2} \epsilon_{ij}
\ee
leaving the other commutators (80) and (82) unchanged.

\noindent
Such a noncommutative space has been studied recently by Lukierski, Zakrzewski and the present author [23] in connection with Galilean symmetry in $(2+1)$-dimensions including a second central charge of the extended algebra. By means of the commutation relations (80), (82) and (83) the Hamiltonian
\be
H_2 = \frac{p_i^2}{2} + \frac{\omega^2}{2} x^2_i
\ee
leads to the EOM in phase space
\begin{eqnarray}
\dot{x}_i &=& p_i + \kappa \epsilon_{ij} x_j\\
\dot{p}_i &=& - \omega^2 x_i
\end{eqnarray}
and, by combining these, we arrive at the EOM (76) again.

\noindent
It is an essential point of the latter approach, that the interaction with the external $B$-field has been shifted from the Hamiltonian to the commutator (83) inducing a noncommutative geometric structure.

\bigskip
As a second alternative we write the Hamiltonian again in the form (84)
\be
H_3 = \frac{u^2_i}{2} + \frac{\omega^2}{2} x_i^2
\ee
where the $u_i$ are the velocities now related to the canonical momenta $p_i$ as usual
\be
u_i = p_i - \kappa A_i
\ee
and the vector potential $A_i$ describes a $B$-field of unit strength
\be
A_i = - \frac{1}{2} \epsilon_{ij} x_j\ .
\ee
With (87) and (88) $H_3$ is identical with $H_1$, but we consider the commutation relations of the velocities as the primary objects now. We obtain for them
\be
\left[ u_i, u_j\right] = i \epsilon_{ij} \kappa
\ee
i.e., we have a noncommutative structure of velocity space now.

\noindent
This approach may be generalized. As (90) is independent of the potential term in (87), we may consider the potential free case. With new variables 
\be
b : = \frac{1}{\sqrt{2}} (u_1 + i u_2), \qquad b^\dagger : = \frac{1}{\sqrt{2}} (u_1 - i u_2)
\ee
the commutator (90) together with the Hamiltonian leads, as is well known, to the oscillator algebra (we put $\kappa = 1$)
\begin{eqnarray}
\left[ b, b^\dagger\right] &=& 1 \\
\left[ H,b \right] &=& - b, \qquad \left[ H, b^\dagger\right] = b^\dagger\ .
\end{eqnarray}
Now we may again consider an arbitrary, nonlinear deformation 
\begin{eqnarray}
\left[ b, b^\dagger \right] &=& \varphi (K) \\
\left[ K,b \right] &=& - b, \qquad \left[ K, b^\dagger \right] = b^\dagger\ .
\end{eqnarray}
For a discussion of this algebra we refer to Sec.~2.2 of this paper. The particular case of a $SU_q (2)$ deformation has been considered by Hojman recently [24].

\section{The ${\bf s}$-equivalence for spherically symmetric potentials revisited}

For a particle moving in an arbitrary spherically symmetric potential $V(r)$ a whole set of dynamically equivalent Hamiltonian structures exists [25], [26]. This set is characterized by the symplectic structure
\be
\{ x_i, u_j\}_\omega = \delta_{ij} - \frac{G L_iL_j}{1+G {\bf L}^2}
\ee
where $L_i$ denotes the $i$-th component of the angular momentum and $G({\bf L})$ is a homogeneous function of degree $(-3)$.

\medskip
For the particular case
\be
G({\bf L}) = \frac{\gamma}{L^3}, \quad \gamma \in {\bf R}^1
\ee
with
\[
L : = \sqrt{{\bf L}^2}
\]
the new Hamiltonian $\tilde{H}$ may be expressed explicitly in terms of canonical variables $({\bf x}, {\bf p})$ as follows [25]
\be
\tilde{H} = \frac{p^2}{2} + \frac{1}{2r^2} (\gamma^2 - 2\gamma J) + V(r)
\ee
where ${\bf J}$ is the canonical angular momentum and $J : = \sqrt{{\bf J}^2}$. 

\medskip
\noindent
If quantized, the spectrum of the Hamilton operator (98) differs from the standard one due to the additional second term. But in quantum mechanics, the EOM derived from (98) is not $s$-equivalent to the standard form
\be
\dot{u}_i = - \frac{x_i}{r} V^\prime (r)\ .
\ee
Let us show this by an explicit calculation:

\smallskip
\noindent
We obtain by means of (98)
\be
u_i : = \dot{x}_i = p_i - \frac{i\gamma}{\hbar r^2} [J, x_i]
\ee
and therefore
\be
\dot{u}_i = - \frac{x_i}{r} V^\prime (r) + \gamma A_{1,i} + \gamma^2 A_{2,i}
\ee
with the quantum corrections $A_i$ defined by
\begin{eqnarray}
A_{1,i} : &=& - \frac{i}{\hbar} [\frac{J}{r^2}, p_i ] - \frac{1}{2\hbar^2} [J, [\frac{x_i}{r^2}, p^2]]\\
A_{2,i} : &=& \frac{1}{r^4} (x_i - \frac{1}{\hbar^2} [J, [J, x_i]])\ .
\end{eqnarray}
If we dequantize these $A_{j,i}$ by means of Dirac's rule we obtain zero as required.
It is easy to show that $A_{2,i}$ e.g. is nonvanishing. By taking matrix elements of $A_{2,i}$ between angular momentum eigenstates, we obtain
\be
< \ell + 1 | A_{2,i} | \ell > = <\ell +1 | \frac{x_i}{r^4} | \ell > B (\ell)
\ee
with
\be
B(\ell) : = 1 - \left( \sqrt{(\ell +1)(\ell+2)} - \sqrt{\ell(\ell+1)}\right)^2\ .
\ee
For finite $\ell$ we have a nonvanishing $B(\ell)$ but it vanishes for large $\ell$ (classical limit) as $0(\ell^{-1})$ in agreement with dequantization.

\section{Interactions as modified commutators in quantum field theory}

In Sec.~3 we observed that for a simple example the interaction can be expressed either in terms of a Hamiltonian or in terms of modified commutators.

\noindent
The application of this idea to quantum field theory is a highly exciting matter. At present we are far away from a systematic treatment of such an idea. In this section it is our aim to give a brief account of some existing examples pointing in this direction. As a matter of convenience we will limit ourselves to nonrelativistic Fermi systems as they appear in solid state physics.

\noindent
A well known example of an integrable model is the $1d$-Luttinger model [27]. It has been shown by Komori and Wadati [28] that this model can be expressed equivalently by two Fermi fields $\psi_j$ ($j = 1,2$) satisfying free field EOM but anyon-like commutation relations for $j \not= k$
\begin{eqnarray}
\psi_j (x) \psi_k^\dagger (y) + \exp (i(-1)^j \lambda) \psi^\dagger_k (y) \psi_j (x) &=& 0\nonumber\\
& & \\
\psi_j (x) \psi_k  (y) + \exp (i(-1)^{j+1} \lambda) \psi_k (y) \psi_j (x) &=& 0\nonumber
\end{eqnarray}
where $\lambda$ is proportional to the coupling strength between the two fields.

\noindent
A quite similar situation arises in two space dimensions if a charged matter field couples minimally to an abelian gauge field described by a Chern-Simons term. This coupling can be removed by a gauge transformation such, that the new matter field will be described by a free Hamiltonian but anyonic commutation relations [29].  This theory is of importance in relation to the fractional quantum Hall effect.

\noindent
In a recent paper P.W.~Anderson et al.~[30] described $2d$-Fermions by first bosonizing them and then modifying the bosonic commutation relations.They didn't succeed in finding the corresponding Fermion representation. But this can be achieved for electrons with an on-site repulsive interaction of infinite strength in the Hubbard model [31]. In this case at most one electron can occupy a lattice site $i$. Usually this will be achieved by means of the Gutzwiller projector
\be
C^\dagger_{i,\sigma} \to C^{\prime\dagger}_{i\sigma} : = (1-N_{i,-\sigma}) C^\dagger_{\sigma,i}
\ee
where now the $C^\prime_{i,\sigma}$ obey complicated commutation relations. But it is easier to introduce field operators satisfying a new on-site algebra
\begin{eqnarray}
d_{i,\sigma} d_{i,\sigma^\prime} &=& 0 \nonumber \\
d_{i,\sigma} d^\dagger_{i,\sigma^\prime} &=& \delta_{\sigma \sigma^\prime} (1-N_{i,-\sigma})\ .
\eea
It may easily be seen, that by operating on state vectors the algebra of the $C^\prime_{i,\sigma}$ or $d_{i,\sigma}$ respectively are the same. But (108) has the advantage to hold as an operator relation.

\medskip
\noindent
The algebra (108) or modifications of it may be generalized to the continuum without any difficulty.

\medskip
It is an open question how to formulate on-site interaction of finite strength in terms of a new Fermionic algebra. For Bosons a first step in this direction has been done quite recently by Flores [32].

\section{Conclusions}

We have seen from the examples given for one- and two-space dimensions, that our question ``how to quantize dynamically equivalent Hamiltonian structures" has presumably a unique answer in those cases where we have a underlying dynamical symmetry. It remains to be shown that this is also true for the $2d$-Coulomb problem as well as for $3d$-examples (hydrogen atom, harmonic oscillator).
A general framework is also missing.

\noindent
The example of a noncommutative $2d$-space inducing the interaction of an oscillator with a constant magnetic fields also calls for generalizations.

\noindent
Finally, in field theory, we have to answer the question in which cases the interaction can be described by nonstandard commutation relations instead of an interaction term in the Hamiltonian.

\section{References}

\begin{description}
\item{[1]} D.G.~Currie and E.J.~Saletan, J.~Math.~Phys. \underline{\bf 7}, 967 (1966)\\
J.~Cislo, J.~Lopuszanski and P.C. Stichel, Fortschr. Phys. \underline{\bf 46}, 45 (1998).

\item{[2]} J.~Douglas,, Trans. Am. Math. Soc. {\bf 50}, 71 (1941).

\item{[3]} E.P.~Wigner, Phys. Rev.~{\bf 77}, 711 (1950).

\item{[4a]} T.D.~Palev, J.~Math.~Phys., {\bf 23}, 1778 (1982). 

\item{[4b]} S.~Chaturvedi and V. Srinivasan, Phys. Rev. {\bf A44}, 8024 (1991).

\item{[5]} P. Okubo, Phys. Rev. {\bf D22}, 919 (1980).

\item{[6]} V.I. Man'ko, G. Marmo, E.C.G. Sudarshan and F. Zaccaria, Int. J. Mod. Phys. (quant-ph/9612007).

\item{[7]} cp. M.A. Lledo and M. Gracia-Sucre, J. Math. Phys. {\bf 37}, 160 (1996);
M.J. Gotay, Obstructions to quantization, preprint math-ph/9809011
and the literature quoted there. 

\item{[8]} J.F. Carinena, J. Clemente-Gallardo, E. Follana, J.M. Gracia-Bondia, A. Rivero and J.C. Varilly, Connes tangent groupoid and deformation quantization, preprint math.DG/9802102.

\item{[9]} R.F. Werner, The classical limit of quantum theory, preprint quant-th/9504016.

\item{[10]} M.A. Rieffel, Questions on Quantization, preprint quant-ph/9712009 v 2.

\item{[11]} C. Leubner and M.A.M. Marte, Phys. Lett. {\bf 101A}, 179 (1984).

\item{[12]} V.I. Man'ko, G. Marmo, E.C.G. Sudarshan and F. Zaccaria, Phys. Scripta {\bf 55}, 528 (1997).

\item{[13]} J. Katriel and C. Quesne, J. Math. Phys. {\bf 37}, 1650 (1996).

\item{[14]} A. Dimakis and F. M\"uller-Hoissen, Some aspects of noncommutative geometry and physics, preprint physics/9712004.

\item{[15]} M. Arik and D.D. Coon, J. Math. Phys. {\bf 17}, 524 (1976).

\item{[16]} L.C. Biedenharn, J. Phys. {\bf A22}, L873 (1989)\\
A.J. Macfarlane, J. Phys. {\bf A23}, 4581 (1989).

\item{[17]} B. Quesne and N. Vansteenkiste, Helv. Phys. Acta {\bf 69}, 141 (1996);
\\ A. Guichardet, J. Math. Phys. {\bf 39}, 4965 (1998).

\item{[18]} M. Ro{\v{c}}ek, Phys. Lett. {\bf B255}, 554 (1991).

\item{[19]} Sicong Jing, J. Phys. A. Math. Gen. {\bf 31}, 6347 (1998).

\item{[20]} R. Finkelstein and E. Marcus, J. Math. Phys. {\bf 36}, 2652 (1995).

\item{[21]} A. Kempf, Phys. Rev. {\bf D55}, 7909 (1997) and J. Math. Phys. {\bf 35}, 4483 (1994).

\item{[22]} B.F. Samsonov, J. Math. Phys. {\bf 39}, 967 (1998).

\item{[23]} J. Lukierski, P.C. Stichel and W. Zakrzewski, Ann. Phys. {\bf 260}, 224 (1997).

\item{[24]} P.A. Hojman, J. Phys. A. Math. Gen. {\bf 24}, 249 (1991).

\item{[25]} M. Henneaux and L.C. Shepley, J. Math. Phys. {\bf 23}, 2101 (1982).

\item{[26]} J. Cislo, J. Lopuszanski and P.C. Stichel, Fortschr. Phys. {\bf 43}, 745 (1995).

\item{[27]} J.M. Luttinger, J. Math. Phys. {\bf 4}, 1154 (1963).

\item{[28]} Y. Komori and M. Wadati, Phys. Lett. {\bf A218}, 42 (1996).

\item{[29]} R. Jackiw and S.Y. Pi, Phys. Rev. {\bf D42}, 3500 (1990),\\
A. Lerda, Anyons-quantum mechanics of particles with fractional statistics, Lecture Notes in Physics m14, Springer 1992.

\item{[30]} P.W. Anderson and D. Khveshchenko, Phys. Rev. {\bf B52}, 16415 (1995).

\item{[31]} J. Hubbard, Proc. Roy. Soc. {\bf A276}, 238 (1963).

\item{[32]} J.C. Flores, Bose-Hubbard Hamiltonian from generalized commutation relations, preprint cond-mat/9712006.
\end{description}

\end{document}